# MID-Radio Telescope, Single Pixel Feed Packages for the Square Kilometre Array: An Overview

A. PELLEGRINI[1], MEMBER, IEEE, J. FLYGARE [2], STUDENT MEMBER, IEEE, I. P. THERON[3], MEMBER, IEEE, R. LEHMENSIEK[3], SENIOR MEMBER, IEEE, A. PEENS-HOUGH[4], J. LEECH[5], M. E. JONES[5], A. C. TAYLOR[5], R. E. J. WATKINS[5], L. LIU[5], A. HECTOR[5], B. DU[6] AND Y. WU[6]

[1]SKA Organization, Lower Withington, SK11 9FT, Jodrell Bank Cheshire, UK
[2]Onsala Space Observatory, Dept. of Space, Earth and Environment, Chalmers University of Technology, SE-41296, Gothenburg, Sweden
[3]EMSS Antennas, 18 Techno Ave, Technopark, Stellenbosch, South Africa
[4]South African Radio Astronomy Observatory (SARAO), Systems Engineering Dept., Cape Town, South Africa
[5]Dept. of Physics (Astrophysics), University of Oxford, Oxford, UK
[6]Joint Laboratory for Radio Astronomy Technology/The 54th Research Institute of CETC, Shijiazhuang, Hebei 050081, China

*(Invited Paper)*

CORRESPONDING AUTHOR: A. Pellegrini (e-mail: a.pellegrini@skatelescope.org).

**ABSTRACT** The Square Kilometre Array (SKA) project is an international effort to build the world's largest radio telescope, enabling science with unprecedented detail and survey speed. The project spans over a decade and is now at a mature stage, ready to enter the construction and integration phase. In the fully deployed state, the MID-Telescope consists of a 150-km diameter array of offset Gregorian antennas installed in the radio quiet zone of the Karoo desert (South Africa). Each antenna is equipped with three feed packages, that are precision positioned in the sub-reflector focus by a feed indexer platform. The total observational bandwidth (0.35-15.4GHz) is segmented into seven bands. Band 1 (0.35 – 1.05 GHz) and Band 2 (0.95 – 1.76 GHz) are implemented as individual feed packages. The remaining five bands (Bands 3, 4, 5a, 5b, and 6) are combined in a single feed package. Initially only Band 5a (4.6 – 8.5 GHz) and Band 5b (8.3 – 15.4 GHz) will be installed. This paper provides an overview of recent progress on design, test and integration of each feed package as well as project and science goals, timeline and path to construction.

## I. INTRODUCTION

The Square Kilometre Array is an ambitious international project to build the world's largest radio astronomy telescope that will enable breakthrough science and discoveries not achievable with the current facilities.

The SKA technology has been demonstrated by precursor telescopes: MeerKAT [1]-[2], built in the Karoo desert (South Africa) and operational since 2018, currently the world's most sensitive telescope array in L-band. ASKAP [3], able to achieve extremely high survey speed, and MWA [4], a low frequency telescope for large surveys of the southern hemisphere, are both commissioned at the Murchison Radio-astronomy Observatory (MRO) in Western Australia. LOFAR [5], commissioned by ASTRON and deployed in The Netherlands with extensions to other European countries, operates at the lowest frequencies observable from Earth.

After more than a decade of groundwork, the pre-construction phase started in 2012 and concluded in December 2019 with a successful System Critical Design Review (SCDR) that established the foundation for the Design Baseline. The baseline design describes one observatory placed in three host countries: the Headquarters (HQ) in UK, the MID telescope in South Africa and the LOW telescope in western Australia [6]-[7]. The combined sensitivity of the two telescopes covers a potentially continuous frequency band that spans from 50 MHz up to 15.4 GHz with provision to extend it up to 26 GHz. The SKA1-LOW [8] consists of an array of ~130,000 dual-polarised, log-periodic antenna elements deployed in the Murchison Shire of Western Australia. The SKA1-MID [9] consists of a total of 197 Gregorian antennas deployed in the Karoo region of the Northern Cape Province of South Africa. The SKA1-MID telescope will be able to achieve high sensitivity over a wide frequency range. To this end, the Single Pixel Feed (SPF) – so called as it creates a single beam per antenna, in one frequency band at a time – sub-element has been divided in the seven bands shown in Table 1. Each feed is precision positioned in the secondary focus by a feed indexer platform. The two lower bands (Band 1 and Band 2) are contained in separate feed packages. They have both successfully passed the Critical Design Review (CDR) and the qualification tests and are currently installed on the antenna prototype for the on-dish testing activities. The five high frequency bands share a single feed package which can be populated in stages. The cryostat and the two initial bands (SPF Band 5a and 5b) are completing the development and test phase and expected to pass CDR in 2021.

**TABLE 1.** Frequency range and fractional bandwidth of the SKA1-MID telescope

| Band | Frequency (GHz) | Fractional Bandwidth |
| --- | --- | --- |
| 1 | 0.35 – 1.05 | 3.00:1 |
| 2 | 0.95 – 1.76 | 1.85:1 |
| 3 | 1.65 – 3.05 | 1.85:1 |
| 4 | 2.80 – 5.18 | 1.85:1 |
| 5a | 4.60 – 8.50 | 1.85:1 |
| 5b | 8.30 – 15.40 | 1.85:1 |
| 6 | 15.00 – 24.00 | 1.60:1 |

Further details on the technology developments for the Single Pixel Feed are provided in this paper along with the status of the project, the timeline and the roadmap to construction.

## II. PROJECT DRIVERS AND EVOLUTION

The proposal for the SKA arose from the demand of the scientific community for new capabilities to address fundamental questions in astronomy.

### A. SCIENCE GOALS

Beginning 2012 eight Science Working Groups (SWGs) developed the science analysis and goals [10]-[11] for the SKA and set the High Priority Science Objectives (HPSO). These were later reviewed in an open process to reassess and refine the key areas of science and eventually define science requirements. Fig. 1 shows the high and medium priority SKA1 observational categories over the available observing frequency bands of the LOW and MID telescopes.
Band 1, Band 2 and Band 5 are the highest priorities for the initial deployment of the MID telescope. Band 3, Band 4 and Band 6 will not be treated in this paper. Nevertheless, the mechanical interfaces and the thermal and electronics infrastructure needed to support them forms part of the present design. A summary of the frequency range covered by the SKA1-MID feeds is shown in Table 1.

### B. TIMELINE

The key dates of the project are shown in Fig. 2. Prototypes of SKA1-MID and -LOW telescopes have been deployed at the host sites and the construction phase is expected to start in 2021, further details are provided in Section C.
It is worth mentioning that in 2019 the SKA Observatory Convention was signed by funding member countries; the SKA Observatory will become an Intergovernmental Organisation (IGO) in full force in early 2021.

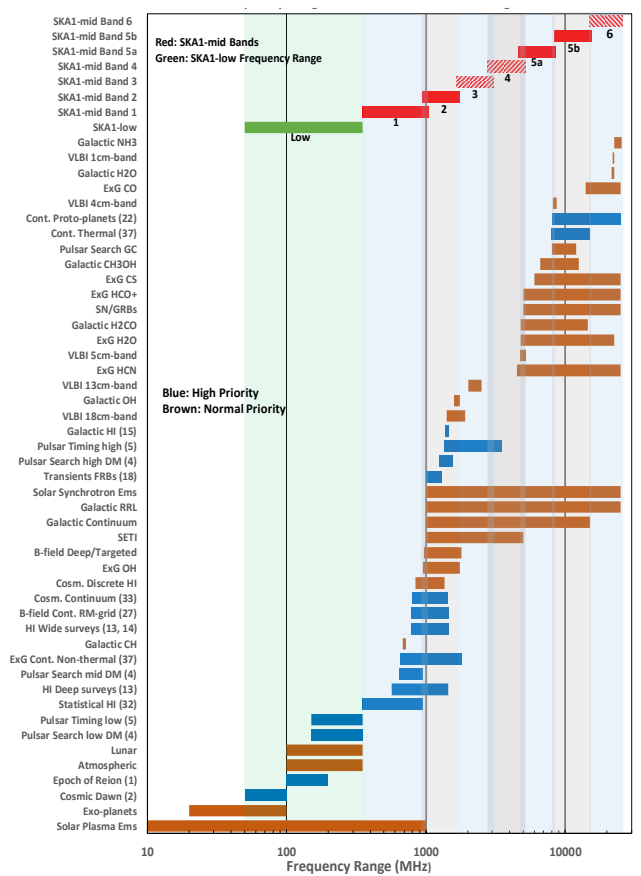

**FIGURE 1.** SKA1 Observational Categories, science priorities and allocation to the electromagnetic spectrum.

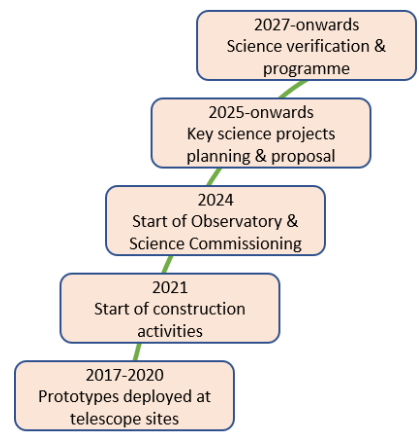

**FIGURE 2.** SKA Project key dates

### C. ROADMAP TO CONSTRUCTION

The central HQ, located in the iconic site of Jodrell Bank, is leading the construction of the observatory. The roll-out-plan for the construction of the SKA1-MID and -LOW telescopes is shown in Fig. 3.
At the time of writing, project management and engineering teams in the SKA Organisation (SKAO) are preparing the contractual data package to start construction in mid-2021.

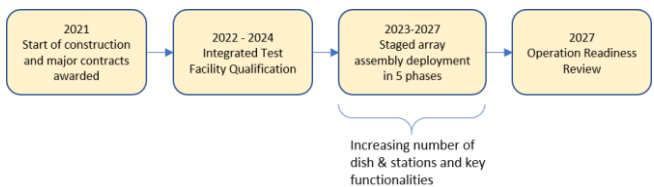

**FIGURE 3.** High level of SKA1 roll-out-plan for construction

Following the start of the contract tendering process and construction activities, a sequential assembly of hardware, software and firmware is rolled out in five phases. The Integration Test Facility (ITF) tests as much as possible of the system in a laboratory environment. The system integration and verification on site is staged in five phases known as Array Assemblies (AA). Each Array Assembly is characterized by a number of dishes (SKA1-MID) or stations (SKA1-LOW), the key functionalities of the array and the resulting scientific capability. The activities end with a demonstration of operational readiness. Science verification and commissioning are tied to the roll-out plans, as they depend on these capabilities.

## III. SKA1-MID TELESCOPE

The MID telescope project is a global endeavour that sees the collaboration of more than 11 countries spread in 9 time zones. Once fully deployed, the SKA1-MID telescope will consist of a combination of 133 15 m shaped Gregorian offset reflector antennas and 64 13.5 m similar, but unshaped dishes from the MeerKAT telescope. The antennas will be located in a 3-arm spiral configuration departing from an inner dense core of 1 km diameter towards the outer part of the array and will extend over a 150 km diameter (Fig. 4) with a total collecting area of ~ 32600 m$^2$.

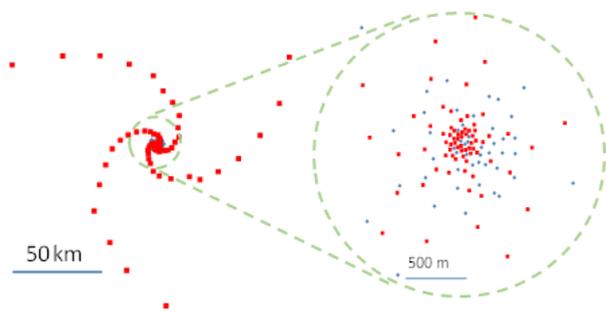

**FIGURE 4.** SKA1-MID array configuration and internal core

The sensitivity ($A_e/T_{sys}$) and the efficiency as well as the feed and receiver noise of the single SKA dish are shown in Table 2. For Band 1 and Band 2 the data have been obtained around the mid-band frequency at an elevation of 45°. The data for Band 5 are instead averaged across the band with the dish pointing at zenith.
The sensitivity and the System Equivalent Flux Density (SEFD) of the whole SKA1-MID array (combination of SKA and MeerKAT dishes) are shown in Table 3 for the four high priority bands.

**TABLE 2.** Single SKA dish sensitivity, efficiency and noise contributions

| Parameter | Band 1 | Band 2 | Band 5a | Band 5b |
|---|---|---|---|---|
| Frequency (GHz) | 0.7 | 1.4 | Average | Average |
| Cosmic Background (K) | 2.73 | 2.73 | 2.73 | 2.73 |
| Galactic (K) | 4.53 | 0.67 | 0.01 | 0.002 |
| Atmosphere (K) | 2.2 | 2.6 | 2.0 [a] | 4.3 [a] |
| Spillover (K) | 3 | 0.9 | 0.8 [a] | 0.2 [a] |
| Total $T_{antenna}$ (K) | 12.5 | 6.9 | 5.5 [a] | 7.2 [a] |
| $T_{receiver}$ (K) [b] | 13.5 | 5.6 | 7.4 | 9.2 |
| $T_{structure\ and\ backend}$ (K) | 1 | 1 | 1.6 | 1.6 |
| $T_{System}$ (K) | 27 | 13.5 | 14.5 [a] | 18.0 [a] |
| Aperture Efficiency, η [c] | 0.81 | 0.90 | 0.84 | 0.83 |
| Effective area, $A_e$, (m$^2$) | 143.1 | 159 | 148.4 | 146.7 |
| $A_e/T_{sys}$ (m$^2$/K) | 5.3 | 11.8 | 10.2 [a] | 8.1 [a] |

a Referred to zenith
b Includes the feed and vacuum window contributions
c Assuming perfect optics, i.e. excluding mechanical tolerance of the structure

**TABLE 3.** SKA1-MID Array Performance

| Frequency band | SEFD (Jy) | Equivalent $A_e/T_{sys}$ (m$^2$/K) |
|---|---|---|
| Band 1 + UHF Band | 2.85 | 967 |
| Band 2 + L-band | 1.55 | 1784 |
| Band 5a | 2.38 | 1161 (2.5 GHz max. sampled band) |
| Band 5b | 2.77 | 998 (2.5 GHz max. sampled band) |

## IV. SINGLE PIXEL FEEDs: OVERVIEW

High antenna efficiency and low total noise are the key drivers of the SPFs design in the individual bands to achieve maximum on-dish sensitivity. While the dish has been designed to accommodate up to seven feeds, at the time this paper is released, four of them are at a mature stage of prototyping and will be discussed in the next sections.

### A. SPF BAND 1

The Band 1 SPF package (shown in Fig. 5), developed at Onsala Space Observatory (OSO) is a room temperature system that operates over the frequency from 0.35 to 1.05 GHz [12]. The SPF is based on a dual linear polarized Quad Ridged Feed Horn (QRFH) of 1.3 m diameter and overall feed package length of 1.5 m.
The QRFH assembly is comprised of three main parts: four 'ridges', an outer flare split into four 'quarter sections' and a back-short section. The feed is mounted on a support that also acts as an interface to the indexer on the antenna. The aperture of the horn is protected by a polycarbonate radome, the feed is made moisture proof by sealing all mechanical parts while a desiccator absorbs any moisture leakage. An environmental shield protects the feed package from rain and direct sunlight.

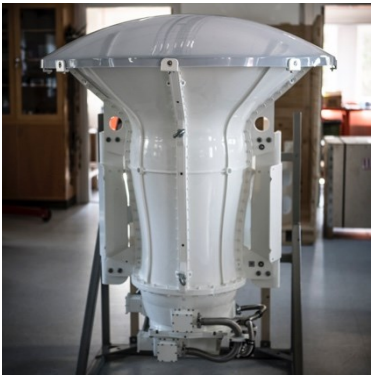

**FIGURE 5.** SPF Band 1 feed package shown without cover shield

The two room temperature low-noise amplifiers (LNA) [13], manufactured by Low Noise Factory (LNF), are integrated in the ridges close to the feed pins of the two orthogonal polarisations, minimizing the losses in front of the LNA. The noise-injection coupler is combined with the LNA in a single assembly with the calibration signal injected prior to the first amplification stage. The rest of the feed electronics (i.e. 2nd stage LNAs, calibration noise diode, monitor and control electronics, etc.) are located in the feed-controller enclosure mounted on the rear of the feed horn. The block diagram of the Band 1 Feed Package is shown in Fig. 6.

The QRFH spline-defined horn and ridge profiles, together with the back short design, are optimized to provide maximum sensitivity over the wide fractional bandwidth (3.0:1) [12]. The phase centre is accurately determined to maximize the phase efficiency and determine the mounting position on the dish feed indexer to achieve maximum sensitivity.

A prototype has been fully qualified to demonstrate compliance with the SKA requirements. Due to the size of the feed package, customized equipment has been developed to carry out some of the qualification tests (e.g. RFI and environment tests). The dish system sensitivity, presented in Fig. 7, has been estimated by combining measurements performed in the laboratory and simulations of the feed on the SKA dish.

The measured receiver noise temperature is below 18 K for both polarisations while the receiver gain is compliant with the SKA requirements as shown in Fig. 8.

Successful tests have been performed with the Band 1 SPF installed on a MeerKAT antenna [14] and on the Dish Verification Antenna 1 (DVA-I) [15], an offset Gregorian SKA telescope precursor, located at the Dominion Radio Astrophysical Observatory (DRAO) in Canada. Although a direct comparison may be not possible, the results of these tests confirmed the simulated system performance of the SKA dish. Installation, integration and verification of the Band 1 SPF on the SKA-MPG dish (funded by the Max Plank Society) is ongoing at the time this paper is released and expected to conclude in 2021.

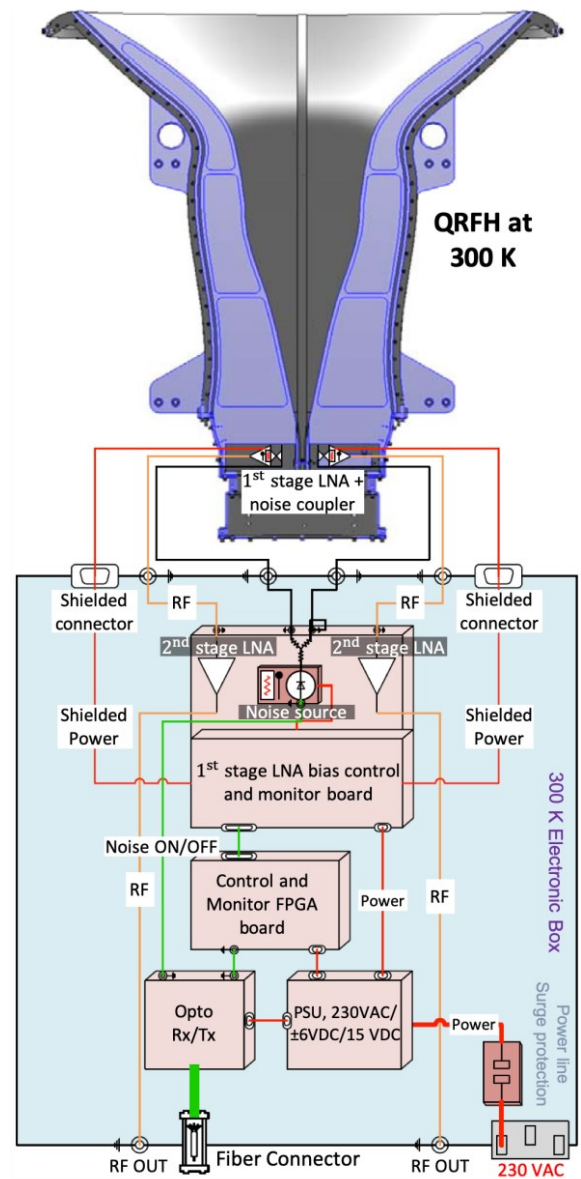

**FIGURE 6.** SPF Band 1 system block diagram

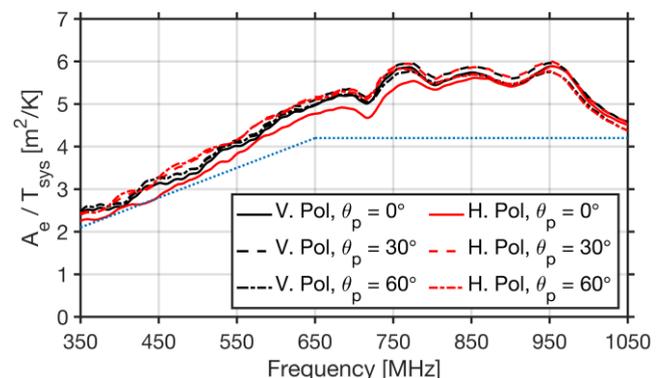

**FIGURE 7.** Sensitivity derived for the H/V polarisations across the Band 1 frequency range for three zenith angles

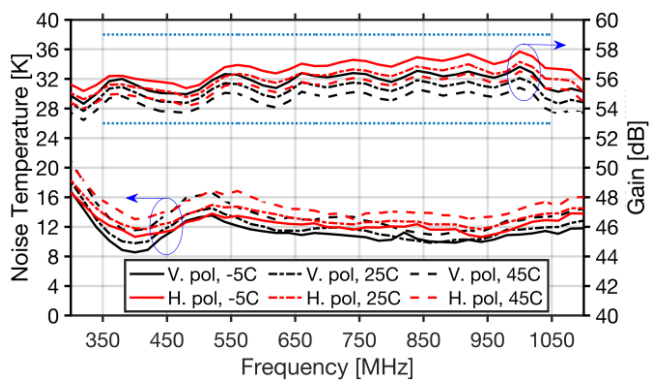

**FIGURE 8.** Receiver gain and noise temperature for the H/V polarisations across the Band 1 frequency range at different temperatures

*SPF BAND 2*

The Band 2 SPF (shown in Fig. 9) developed by EMSS Antennas operates over the frequency band 0.95 GHz to 1.76 GHz.

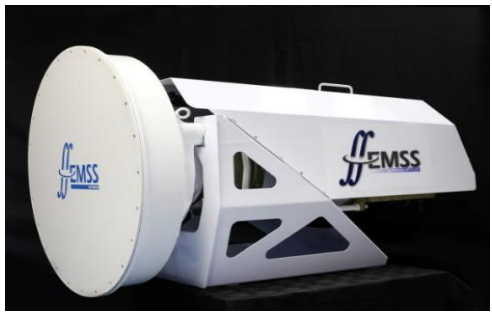

**FIGURE 9.** SPF Band 2 package

The Band 2 SPF block diagram is shown in Fig. 10; the feed package consists of an ambient temperature wide flare angle axially corrugated conical horn [16], a cryogenic Orthogonal Mode Transducer (OMT) realised as a pair of orthogonal dipoles [17], LNAs cooled to below 20 K and a final room temperature amplification and matching stage. Cooling is achieved with a Gifford McMahon (GM) cryogenic-cooler that requires high pressure helium. The cryogenic components inside the cryostat are thermally insulated by a high vacuum. The SPF vacuum service supplies the initial vacuum during start-up, while the cryo-pump maintains vacuum during operations. The waveguide is at ambient temperature with a High Density Poly Ethylene (HDPE) dome over the cryogenic dipoles creating the vacuum window. The calibration noise source is thermally stabilized at ambient temperature inside the cryostat to minimize amplitude variations on the signal path. The feed package is covered with a sun shield to reduce solar heating and provide protection against direct rain. Moisture collection is limited by protecting the horn aperture with a hydrophobic radome membrane and desiccator breather.

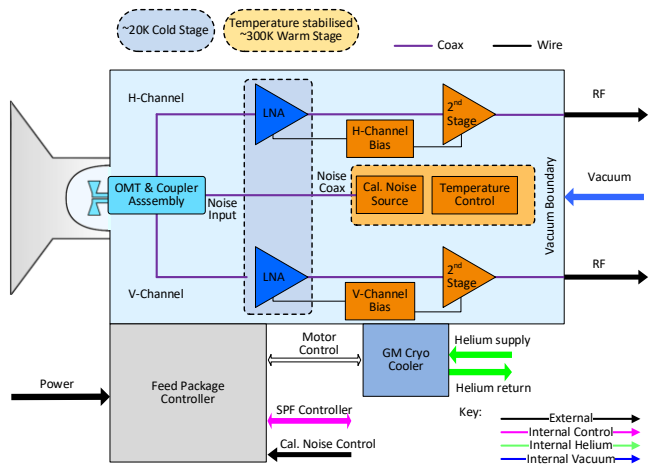

**FIGURE 10.** SPF Band 2 system block diagram

Extensive simulation work has been carried out on the performance analysis of the Band 2 feed on the SKA 1 dish geometry. In [18]-[20] the authors present a method to estimate and maximise the telescope sensitivity over the frequency range allocated to Band 1 and Band 2.

Qualification tests have been performed on the Band 2 SPF prototype and have successfully demonstrated compliance with the SKA requirements. The receiver temperature has been evaluated by using the Y-factor method [21] and the results are shown for both polarisations in Fig. 11. A linear increase from 5 K at the lowest frequency end to 6.5 K at 1760 MHz has been used to evaluate the sensitivity. The system temperature has been calculated as the sum of the receiver temperature, a 1 K contribution allocated to all other effects (i.e. surface, loss, digitizing noise, unexpected effects, etc.) and the antenna temperature.

The latter (shown in Fig. 12) has been calculated using the brightness temperature from [22] with the following parameters:

- $T_{gnd}$ = 300, Ground Temperature
- $eps\_r$ = 3.5, relative permittivity of ground
- $H_a$ = 100, Height of atmosphere
- $T_{go}$ = 20, base temperature
- $f_0$ = 0.408, base frequency
- $T_{cmb}$ = 2.73, Cosmic microwave background emission
- $R_e$ = 6370.95 earth radius [km]

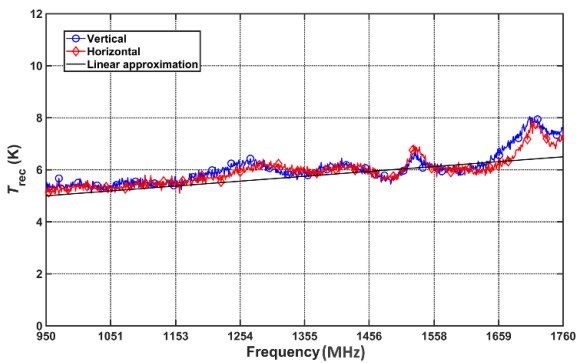

**FIGURE 11.** Measured Band 2 receiver noise temperature for both polarisations

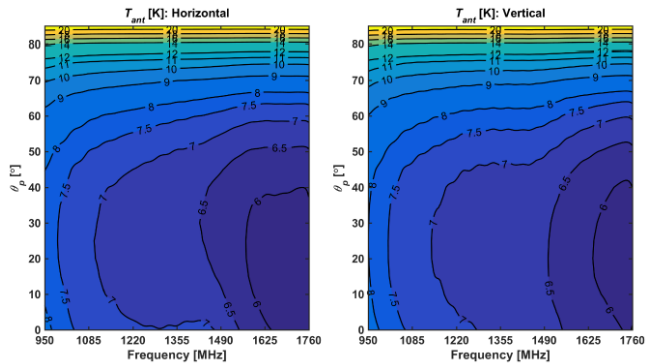

**FIGURE 12.** Antenna temperature as function of frequency and zenith angles for both polarisations

The dish system sensitivity is shown in Fig. 13 for several zenith angles.

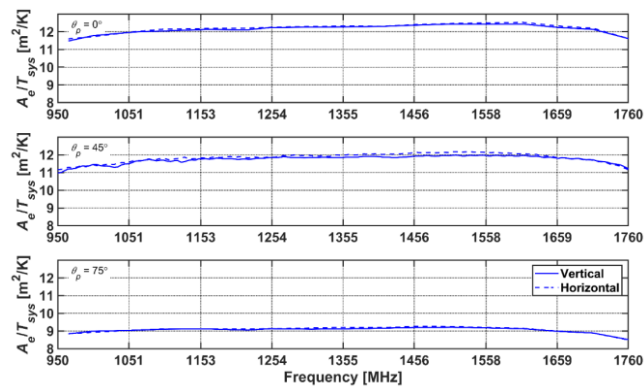

**FIGURE 13.** Sensitivity derived for the H/V polarisations across Band 2 frequency range for three zenith angles

### SPF BAND 345(6)

The SPF Band 345(6) feed package has been developed at the University of Oxford, Department of Physics. It is designed to house up to five feed horns for Bands 3, 4, 5a, 5b and 6 in a single, modular cryostat cooled by a single Gifford McMahon (GM) coldhead. This design was adopted because constraints on the electrical power available for cryogenics at each antenna (5 kW) made running each band from a separate coldhead impractical. By careful cryogenic design, a single Oxford Cryosystems Coolstar 6/30 coldhead which has 30 W of first stage cooling (@ 80K) and 6 W of second stage cooling (@ 10 K) is expected to provide sufficient cooling for up to 5 bands. In the first instance, the higher science priority band 5a and 5b feed horns, OMTs and RF chains will be populated, with the intention of adding bands 3, 4 and possibly 6 at a later date. Fig. 14 shows the Band 345(6) feed package, with bands 5a and 5b populated, mounted on the feed indexer (centre) alongside the separate Band 1 and Band 2 feed packages.

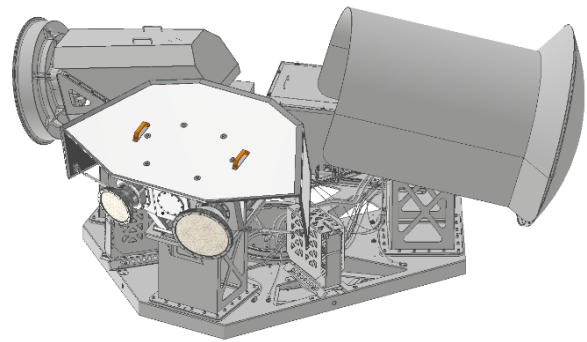

**FIGURE 14.** The Band 345(6) feed package with bands 5a and 5b feed fitted (centre) mounted on the feed indexer. Also visible is SPF Band 1 (right) and SPF Band 2 (left).

Fig. 15 shows the Band 345(6) feed package. Visible at the front are the cryostat windows for the band 5a and 5b feed horns. At the rear of the cryostat is an RFI-tight enclosure for the Feed Package Controller (FPC), which houses the unit's power supplies and monitoring and control electronics. Mounted on the right is an RFI-tight enclosure for a turbomolecular vacuum pump. The unit is protected by a lightweight weather/sun shield and is mounted on a welded steel support frame. The feed package with the full complement of receiver bands is also illustrated.

The cooling power from the GM coldhead (mounted vertically towards the centre of the cryostat) is directed to the feed horns, OMTs and LNAs via copper "thermal plumbing" (Fig. 16). The feedhorns are cooled to the coldhead's first stage temperature of ~80 K, whereas the OMTs and LNAs are cooled to the second stage temperature of ~12 K. The feed horns for each of the two bands are 1:1.85 bandwidth wide flare-angle corrugated horns. Each horn is positioned behind a vacuum window consisting of a Mylar sheet mechanically supported by a closed-cell polyethylene foam (Zotefoam) sheet, which also acts as an infrared block. The horns (cooled to 80K) couple the incoming radiation, via a narrow thermal waveguide break, to the OMTs (at 12K) which couple the two vertical and horizontally polarized $TE_{11}$ waveguide modes into standard semi-rigid cable. The Band 5a OMT is a quad-ridged finline design, whereas the higher frequency Band 5b OMT is a waveguide turnstile type. The horns and OMTs were designed and manufactured by JLRAT/CETC54, China [23]-[25].

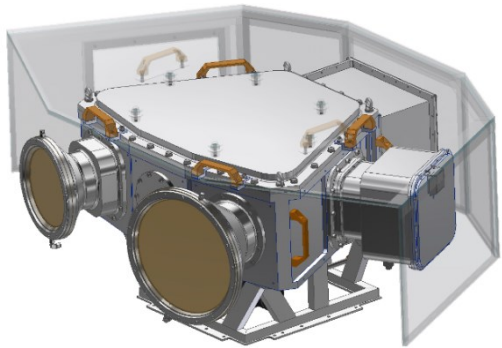

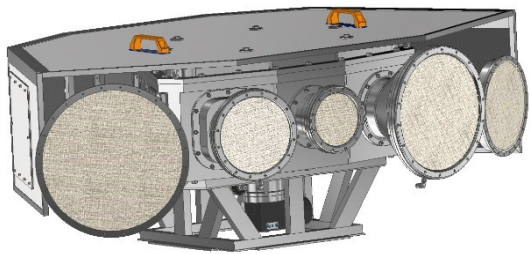

**FIGURE 15.** (Top) The Band 345(6) feed package with bands 5a and 5b cryostat windows and feeds fitted. (Bottom) The Band 345(6) feed package showing the addition of side mounting modules to accommodate the band 3 and 4 feed horns. The band 6 feed horn and window can be seen between mounted centrally between the 5a and 5b feeds.

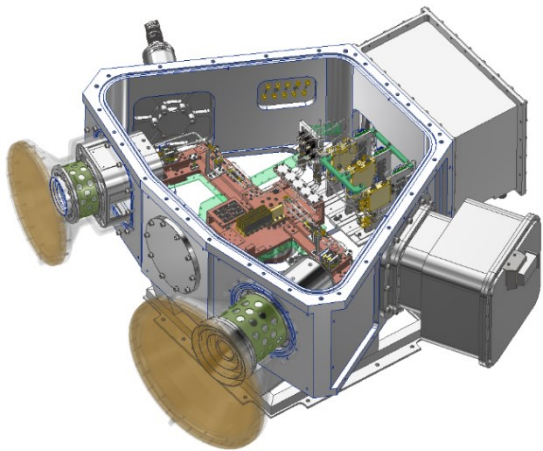

**FIGURE 16.** The Band 345(6) feed package, showing the cryostat interior. Visible is the copper thermal plumbing (with radiation shielding and multilayer insulation removed) and the mounting plates for the warm RF electronics towards the rear of the cryostat. The cryostat windows are shown in transparent brown and the apertures of the corrugated feed horns are visible behind them.

The block diagram of the Bands 345(6) electronics is shown in Fig. 17.

**FIGURE 17.** Block diagram for the Band 5a RF chain. The Low Noise Factory LNA is temperature stabilized by a heater resistor under PI loop control by the Feed Package Controller (FPC). The warm (300K) RF components (warm amplifier, bandpass filter and slope compensating amplifier) are fabricated on a single Duroid substrate housed in a machined aluminium enclosure and temperature stabilised using an analogue PI control circuit.

The orthogonally polarized signals propagating along coaxial cables from each OMT are first amplified by commercial LNAs manufactured by Low Noise Factory. The signals are then directed, via thermally isolating stainless steel/BeCu semi-rigid coaxial cables to a warm RF chain which consists of two additional stages of amplification either side of a band-defining bandpass filter. For each band these components are integrated onto a single planar substrate within a machined aluminium enclosure. These warm RF units are mounted on a 310 K temperature controlled mounting plate inside the cryostat to the rear of the coldhead and thermal plumbing structure. Here we also find a temperature controlled broadband RF noise source for calibration purposes. This noise calibration signal can be RF switched to the band in use and is injected into the throat of each feed horn via a weakly coupled waveguide probe.

The noise performance of the 5a and 5b RF chains have been modelled to obtain estimates for the likely $T_{rec}$ receiver noise performance. The system noise temperatures are expected to be between 8.2 and 9.5 K for band 5a and 9.4 and 13.2 K for band 5b. Electromagnetic modelling of the diffraction between the horn aperture and cryostat window has been performed using CST Microwave Studio. The window has a significant effect on the feed beam pattern compared to an isolated horn, but by careful design of the window aperture and position it is possible to maintain the overall A/T of the antenna (Fig. 18). The full beam patterns for each feed and window on the SKA-MID dish have been calculated using Ticra's GRASP physical optical modelling package. The beam patterns were then used with atmospheric and ground emission models [22] to calculate the expected $T_{spill}$ spillover contribution to the system noise temperature ($T_{sys} = T_{rec} + T_{spill}$). Finally, these system temperature estimates were used to calculate the expected single antenna point source sensitivity ($A_e / T_{sys}$) (Fig. 18 and Fig. 19).

The SPF Band 345(6) prototype cryostat, horns, OMTs and RF components have now been manufactured (Fig. 20) and are currently undergoing assembly, integration and testing at the University of Oxford. Completion of the Critical Design Review and beginning of on-dish testing are expected in 2021.

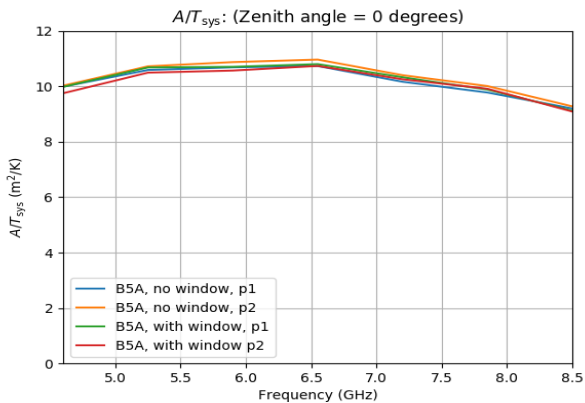

**FIGURE 18.** Sensitivity estimates (for two polarizations p1 = horizontal and p2 = vertical, with and without cryostat windows) for a single antenna pointing at zenith as a function of frequency for Band 5a (4.6 – 8.5 GHz).

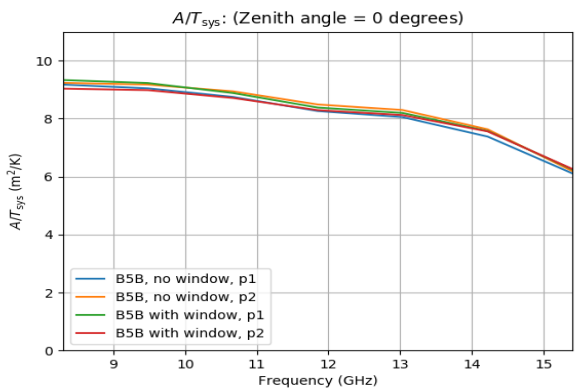

**FIGURE 19.** Sensitivity estimates (for two polarizations p1 = horizontal and p2 = vertical, with and without cryostat windows) for a single antenna pointing at zenith as a function of frequency for Band 5b (8.3 – 15.4 GHz).

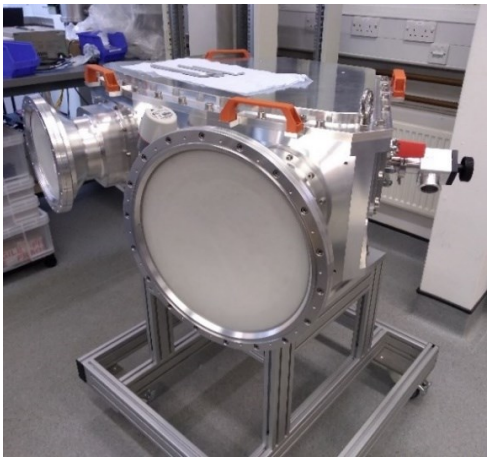

**FIGURE 20.** Band 345(6) prototype cryostat ready for assembly and integration at the University of Oxford.

## IV. CONCLUSION

An overview of the goal, science and road to construction of the Square Kilometre Array radio-telescope has been provided in this paper.

The latest technology developments on the design and prototyping of the Single Pixel Feeds for the SKA1-MID telescope have been discussed. Extensive qualification tests are being carried out at the time this paper is released and will be subject of future publications.

The consolidated effort from all the teams involved demonstrates that the SKA project is mature to enter the construction phase and is certainly on the path to success.

### ACKNOWLEDGMENT

The work detailed in this paper would not have been possible without the collaborative efforts of the international SKA Dish consortium, one of 12 engineering consortia responsible for designing the look and functionality of the SKA's different elements and ensuring they all work together. The consortia brought together hundreds of experts from research institutions and industry around the world, whose talents and inputs were invaluable throughout the SKA's design phase and in preparations for SKA construction.

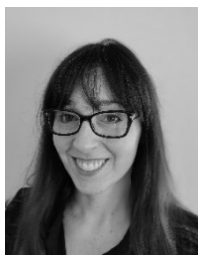

**ALICE PELLEGRINI** was born in Rome, Italy in 1980. She received the Master degree (summa cum laude) in Telecommunication Engineering with specialization in Applied Electromagnetism in October 2005 and, in 2009 a PhD in Information Engineering from University of Pisa, Italy. In 2008 she was awarded a Visiting Scholar position at the Electrical Engineering Dept. of the Penn State University, State College, PA, US. Her main research activities concerned numerical methods for antenna design and propagation with focus on phase arrays. After few years working as Post-Doctoral Research Assistant at Queen Mary University of London, she joined Cobham Antenna Systems as Antenna Design Engineer in 2013. She moved into a System Engineering position at Inmarsat, London, UK in 2016 working on design, verification and commissioning of Earth Station RF systems and antennas for communication and in-orbit testing systems. In 2019 she joined the Square Kilometre Array Organisation based in Jodrell bank, UK as RF and Antenna Domain Specialist. She is responsible for all technical aspects of the design, integration and verification of the antenna and front-end system for the MID-telescope that will be deployed in South Africa. Her scientific interests include radio telescopes, electrically large antennas, satellite communication, antenna design and characterization, and radio frequency subsystems.

Dr. Pellegrini has authored several papers published in international peer-reviewed scientific journals and proceedings of international conferences.

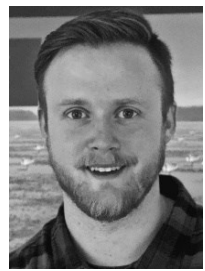

**JONAS FLYGARE** (S'16) was born in Mölndal, Sweden, in 1988. He received the B.Sc. and M.Sc. degrees in engineering physics and the Licentiate degree from the Chalmers University of Technology, Gothenburg, Sweden, in 2016 and 2018, where he is currently pursuing the Ph.D. degree in radio and space science.

In 2012, he was a Test Technician in semi-conductor research and development with CBRITE Inc., Goleta, CA, USA, where he was developing the thin-film-transistor technology for next-generation flat screens. In 2014, he was with the Department of Electrical Engineering, Chalmers University of Technology, where he was working with the gap-waveguide technology. From 2014 to 2016, he was a Project Researcher with the Onsala Space Observatory, Swedish National Facility for Radio Astronomy, Gothenburg, where he was involved to design the feed antenna for the Square Kilometre Array (SKA) Band 1 in radio astronomy. In 2020, he was invited as a visiting student researcher at the Astronomy department at California Institute of Technology, Pasadena, CA, USA where he was working with wideband feed development for the Deep Synoptic Array (DSA). His current research interests include design and system characterization of ultra-wideband antennas and reflector feeds, low-noise amplifiers, and receivers for radio and millimeter wave applications in radio astronomy.

Mr. Flygare received the TICRA Travel Grant by the IEEE APS/URSI, Boston, MA, USA, in 2018.

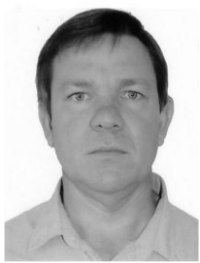

**ISAK P. THERON** (S'88-M'96) was born in Upington, South Africa in 1967. He received the B.Eng. and M.Eng. degrees in electronic engineering, both *cum laude*, and the Ph.D. (Eng.) degree (working on wave propagation in anisotropic chiral media) from the University of Stellenbosch, Stellenbosch, South Africa, in 1989, 1991, and 1995, respectively. He joined EM Software and Systems, Stellenbosch, South Africa, in 1997 working on the computational electromagnetic code FEKO. In 2006, he joined EMSS Antennas, Stellenbosch, South Africa, where he is primarily involved in the design of antennas and cryogenic feed systems for radio astronomy.

Dr. Theron is leading the single pixel feed sub-element development in the dish consortium.

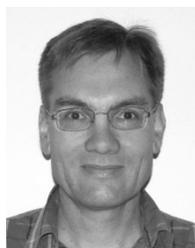

**ROBERT LEHMENSIEK** (S'98–M'01–SM'10) was born in Swakopmund, Namibia. He received the B.Eng, M.Eng, and Ph.D. degrees in electronic engineering from the University of Stellenbosch, Stellenbosch, South Africa, in 1992, 1995, and 2001, respectively.

Since 2003 he has been with EMSS Antennas, Stellenbosch, South Africa, where he was involved in the design of various antennas and microwave components. In the later years, he was mainly involved in the design of the reflector optics and cryogenic receiver systems for the Karoo Array Telescope (KAT), the Square Kilometer Array (SKA), and the Next Generation Very Large Array (ngVLA) radio telescopes.

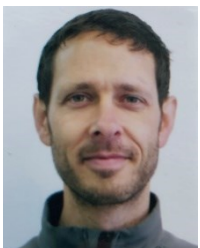

**ADRIAAN PEENS-HOUGH** was born in Pretoria, South Africa in 1978. He received the B.Sc. in Electronics Engineering (*cum laude*) from the University of Pretoria in 2000, and the B.Sc. Hons in Applied Mathematics from the University of Cape Town in 2008. In 2006 he joined the National Research Foundation's project that eventually gave rise to the South African Radio Astronomy Observatory, where he is currently still the primary System Analyst within the Systems Engineering department. He has been responsible for the performance analysis, design and verification of the end-to-end signal paths of the KAT-7 and MeerKAT radio telescopes, and since 2013 in a similar role within the SKA1-MID Dish design consortium.

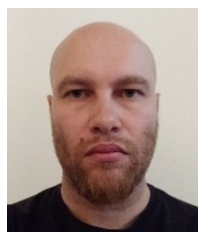

**JAMIE LEECH** was born in Birmingham, U.K, in 1975. He obtained an M.Sci. in natural sciences (experimental and theoretical physics) from Cambridge University, U.K. in 1998 and a Ph.D. in astrophysics, also from Cambridge University, in 2002.

He was a support astronomer at the James Clerk Maxwell Telescope between 2002 and 2006, and is currently a senior researcher at the Department of Physics, University of Oxford, U.K. Current research interests include instrumentation for radio and submillimetre astronomy and radio observations of polarized CMB foregrounds. He is the technical lead of SKA Band 5 SPF development at Oxford.

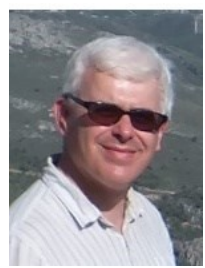

**MICHAEL E. JONES** was born in Wrexham, UK, in 1966 and studied Natural Sciences and Electrical Science at Cambridge University. He obtained a PhD in Radio Astronomy at the Cavendish Laboratory, Cambridge in 1990 and subsequently worked there as a Research Associate and Senior Research Associate. In 2005 he was appointed first as a Lecturer and then as Professor of Experimental Cosmology at the University of Oxford Department of Physics.

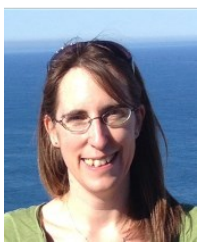

**ANGELA C. TAYLOR** was born in Dartford, UK and studied Natural Sciences at Cambridge University, earning a PhD in Radio Astronomy at the Cavendish Laboratory, Cambridge in 2002. She continued to work in the field of astrophysics and instrumentation as a postdoctoral research assistant at the Cavendish Laboratory until 2005 when she moved to the University of Oxford as a founding member of the Experimental Cosmology Group. She is currently a Professor of Experimental Astrophysics at the University of Oxford working on a range of cosmology and instrumentation projects. Prof. Taylor is a Co-Investigator on the SKA project within Oxford, with specific responsibility for the Band 345 single pixel feed sub-element development in the SKA dish consortium.

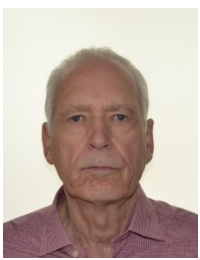

**ROBERT E. J. WATKINS** was born in Cardiff (UK) in1949. He was awarded a BSc 1st Class Honours in Electronic Engineering from the University of Salford in 1971 and a DPhil from the University of Sussex in 1978 for his work on the effects of high dose ion implantation into metals. After a number of post-doctoral positions at AERE Harwell (UK) and the University of Trento (Italy) he took up a position at the Department of Engineering Science, University of Oxford in 1981, and then moved to the Department of Atmospheric, Oceanic and Planetary Physics in 1985 where he worked continuously on test and calibration facilities for space-borne instrumentation. Since 2016 he has been involved with the design and construction of the SPF-B5 cryostat for the Square Kilometre Array, with the work being carried out at the Department of Astrophysics, University of Oxford.

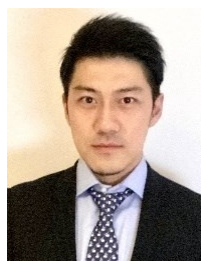

**LEI LIU** was born in ZhengZhou, China, in 1985 and studied Computer Science at Nanjing University of Posts and Communications. He obtained a PhD in Microwave engineering at Northumbria University at Newcastle upon Tyne, UK in 2013. He then worked in the field of astrophysics and instrumentation (development of phased array feed technology) as a Research Associate at Jodrell bank Observatory, the University of Manchester. In 2018 he works as a senior Researcher at Experimental Cosmology group for SKA mid-receiver development at the University of Oxford Department of Physics.

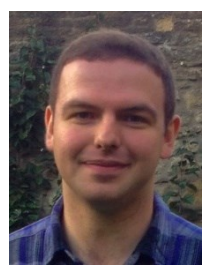

**ANDRE HECTOR** was born in 1991 in L'viv, Ukraine and obtained an M.Sci. in Physics from Royal Holloway, University of London in 2013. From 2013 he studied for a DPhil in millimeter-wave receiver technology at the University of Oxford, specialising in the design and development of cryogenic receivers, feeds and local oscillators. In 2018 he joined the SKA project in Oxford, working as a researcher on the development of the Band 345 system.

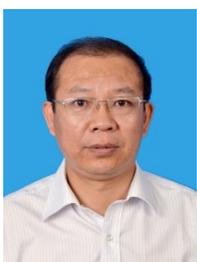

**BIAO DU** was born in Shijiazhuang, China, in 1962. He received the B.S. degree in Electronic Engineering from Xidian University, Xi'an, and the M.S. degree in and Electromagnetic Field and Microwave Technology from the 54th Research Institute of China Electronics Technology Group Corporation (CETC 54), and the Ph.D. degree in Electronic Engineering from Shanghai University, Shanghai, China, in 1983, 1987, 1996, respectively.

From 1987 to 1999, he joined CTEC 54 as an antenna design engineer and senior engineer, Shijiazhuang. From 1999 to 2000, he was a Visiting Scholar with the Department of Electronic Engineering, City University of Hong Kong, Hong Kong. He is currently a Professor, a Chair Expert, and a Deputy Chief Engineer with the CETC54. He is the chief designer of Shanghai 65m Tian Ma radio telescope antenna and Square Kilometre Array Dish Prototype (SKA-P). In 2019, he was appointed as Guest Professor position at Xidian University. He has authored more than 100 academic papers. His research interests are in reflector antennas, corrugated horns, ultra-wideband feeds, microstrip antennas, phased array antennas and metamaterial antennas.

Dr. Du is a Fellow of Chinese Institute of Electronics. He was a recipient of Bao Gang Education Award for the excellent students in 1995. He obtained the special allowance of the State Council of China for his outstanding contribution in natural science in 2016. He also received eight Science and Technology Progress Awards.

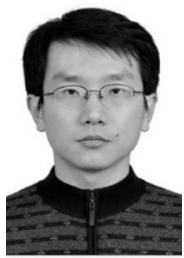

**YANG WU** was born in Shijiazhuang, China in 1984. He received the the Ph.D. degree from the Xidian University, Xi'an, China, in 2013. From September 2014 to Apirl 2017, he worked in National Astronomical Obsevertories, Chinese Academy of Sciencies as a post doctor. Now, he joined the 54th Research Institute of CETC, Shijiazhuang, China, where he is primarily involved in the design of antennas and feed systems for radio astronomy.

Dr. Wu is leading the band 4 and band 5 feed assemblies development of the single pixel feed sub-elment in the dish consortium.